\documentstyle[12pt]{article}

\newcommand{\be}{\begin{equation}}
\newcommand{\ee}{\end{equation}}
\newcommand{\bea}{\begin{eqnarray}}
\newcommand{\eea}{\end{eqnarray}}
\newcommand{\lsim}{{\mathrel{\rlap{\hbox{$<$}}
    \raise4pt\hbox{$\sim$}}}}
\newcommand{\gsim}{{\mathrel{\rlap{\hbox{$>$}}
    \raise4pt\hbox{$\sim$}}}}

\newcommand{\AmS}{{\protect\the\textfont2
  A\kern-.1667em\lower.5ex\hbox{M}\kern-.125emS}}

\begin{document}
\sloppy
\begin{titlepage}
\samepage{
\setcounter{page}{0}
\rightline{CERN-TH.7436/94}
\rightline{hep-ph/9409249}
\vspace{.2in}
\begin{center}
{\bf ANOMALOUS EVOLUTION \\OF NONSINGLET STRUCTURE FUNCTIONS\\}
\vspace{.3in}
{Stefano Forte\footnote{
On leave from INFN, Sezione di Torino, Turin, Italy
}and Richard D. Ball\footnote{On leave from a Royal Society University
Research Fellowship}\\}
\vspace{.15in}
{\it Theory Division,
 CERN\\ CH-1211 Gen\`eve 23, Switzerland\\}
\end{center}
\vspace{.25in}
\begin{abstract}

We review a formalism that includes  the effects of
 nonperturbative U(1) symmetry breaking on the QCD evolution
of nonsinglet structure functions.
We show that a
strong scale dependence  is generated
in an
intermediate energy range $0.5 \lsim Q^2 \lsim 5$~GeV$^2$ for all
values of $x$. We show that this explains naturally the observed
violation of the Gottfried sum, and allows a determination of the shape of
the nonsinglet structure function, in excellent agreement with
experiment.
We argue that these effects may affect the determination of $\alpha_s$
from deep--inelastic scattering.

\end{abstract}
\vspace{.2in}
\begin{center}
{\it Presented at {\bf QCD '94}\\
Montpellier, France, June 1994\\}
\vspace{.1in}
{To be published in the proceedings}
\end{center}
\vfill
\leftline{CERN-TH.7436/94}
\leftline{September 1994}
}
\end{titlepage}
\setcounter{footnote}{0}
Accurate measurements of nonsinglet nucleon structure functions
have been
accomplished only rather recently. Because gluons decouple
from nonsinglet quantites, such measurements provide a direct
 handle on the quark structure of the nucleon, and their scale
dependence can be reliably calculated in perturbative
QCD. In particular, the first moment of the
nonsinglet nucleon structure function
$F_2^{\rm NS}=F_2^p-F_2^n$ has been determined  rather accurately\cite{NMC}:
\be
\label{sgexp}
S_G \equiv \int {dx \over x}\, F_2^{\rm NS}=0.235\pm 0.026.
\ee

This experimental result\footnote{Very
recent data\cite{bade}
show evidence for shadowing in scattering on deuterium. If
this effect is confirmed the value Eq.(\ref{sgexp}) should be reduced by
a further 10--15 \%} is somewhat puzzling in view of the
interpretation of $F_2^{\rm NS}$ and its first moment $S_G$ in the
QCD parton model: in the DIS scheme to all orders
\be
\label{ftpm}
F_2 (x)=x \sum_{i} e_i^2
(q_i (x) + \bar{q}_{i} (x)),
\ee
where $q_i(x)$ are the distribution of quarks of flavor $i$ in the
given target,
so that,using isospin,
\be
\label{ftns}
F_2^{\rm NS}={x\over 3}\left (u^p(x)+\bar u^p(x))-(d^p(x)+\bar d^p(x))\right),
\ee
where $u^p$
 and $d^p$ are up and down quark
distributions in the proton.
Now,  $S_G$ is  to leading order scale
independent; hence at all scales $S_G$ would be expected to be
close to its quark model value, obtained assuming  the
quark distributions in Eq.(\ref{ftns}) to be  given by valence
quarks: this,
however, leads to $S_G={1\over 3}$.

A possible interpretation of this discrepancy  is  that isospin
is broken;
however, isospin violation in this channel is \cite{megott}
at least one order
of magnitude smaller than required to explain Eq.(\ref{sgexp}).
Thus the data (if
correct) imply  that it is the identification of the quark distributions in
Eq.(\ref{ftns}) with valence distributions to fail:
the sea must have a large nonsinglet component, anticorrelated
to the valence. It is easy to reproduce this sort of scenario in
effective models of the nucleon\cite{megott}. However, due to
the (one-loop) scale independence of $S_G$, such a nonsinglet component
cannot be generated perturbatively from the starting valence
distributions.

\bigskip

The evolution equation for nonsinglet quark distributions takes the
simple general form\cite{guido}
\be
{d\over dt}\left[q\pm\bar q\right]=
\left({\cal Q}_{qq}\pm {\cal Q}_{q\bar q}\right)\otimes\left[q\pm\bar q\right],
\label{tlap}
\ee
to any perturbative order, where
\be
\label{qdef}
{ q=u-d; \quad {\cal Q}_{qq}\equiv {\cal P}_{qq}^D-{\cal P}_{qq}^{ND},}
\ee
and ${\cal P}^D$ (${\cal P}^{ND}$) is any quark--quark splitting function
${\cal P}_{q_iq_j}$
such that $i=j$ ($i\not =j$).
At one loop, the only
process which leads to nonsinglet evolution is gluon radiation
(Fig.1a),
which contributes to $P_{qq}^{D}$ but has
vanishing first moment. This conservation holds true to
all orders in the charge-conjugation odd case, but fails beyond
leading order in the C-even one. Two-loop evolution
is then driven by the diagram of Fig.1b: whereas the flavor-invariance of
QCD leads one to expect that this should contribute equally to
${\cal P}^D$  and ${\cal P}^{ND}$, due to Fermi
statistics the final state must be antisymmetrized with respect to the
two identical quarks when $i=j$, while this is not the case if
$i\not=j$. This is enough \cite{rosac} to lead to nonvanishing values
for all
the nonsinglet anomalous
dimensions, which, however, remain very small.

In QCD, axial U(N$_f$) flavor symmetry is broken down to SU(N$_f$)
by nonperturbative effects
due to the axial anomaly. This
leads to a
large difference between flavor-preserving and flavor-changing
transitions\cite{ehq}, and thus
to sizable values of the anomalous dimensions\cite{us}. Indeed, assume
that the emitted and unobserved $q$--$\bar q$ pair may
form a bound state
(Fig.1c). Emission of neutral (charged) mesons will then contribute to
${\cal P}^{D}_{qq}$ (${\cal P}^{ND}_{qq}$). In the pseudoscalar
sector, where U(1) symmetry breaking manifests itself, $\pi^\pm$ emission
will thus contribute to  ${\cal P}^{ND}_{qq}$, and $\pi^0$ and $\eta$
emission to ${\cal P}^{D}_{qq}$ (considering for simplicity the SU(2)
case). Because of the much larger mass of the $\eta$ the flavor-preserving
process is suppressed, thus leading to a large
negative value for ${\cal Q}_{qq}$. This leads to sizable
evolution, with the sign required to explain the result Eq.(\ref{sgexp}).

This argument can be made quantitative \cite{us}
by setting up generalized evolution equations which
include an effective coupling to bound states of the form of Fig.1c.
Because symmetry breaking only appears in the Goldstone sector,
only pseudoscalar mesons need to be included. Introducing a nonsinglet
pion distribution $\pi\equiv \pi^+-\pi^-$, the
nonsinglet evolution equations are then
\bea
{d\over dt} q&=&  {\cal Q}_{qq} \otimes q
+{\cal Q}_{q\bar q}\otimes \bar q+ {\cal P}_{q\pi}\otimes\pi,
\label{qev}\\
{d\over dt} \bar q&=&  {\cal Q}_{qq} \otimes\bar q
+  {\cal Q}_{q\bar q}\otimes q- {\cal P}_{q\pi}\otimes \pi,
\label{qbarev}\\
{d\over dt} \pi&= & {\cal P}_{\pi q} \otimes (q-\bar{q})
+ {\cal P}_{\pi\pi} \otimes\pi
\label{piev}.
\eea
\bigskip

It is still
necessary to specify the effective quark-meson coupling, denoted by a
blob in Fig.1c.
The total cross section for the process of
Fig.1c turns out to depend \cite{us} only on a pseudoscalar and an axial
vertex function; the former dominates  for
intermediate values of $Q^2$ (up to  a few GeV$^2$),  while
the latter controls  the
large $Q^2$ tail.
As a a consequence,
the bulk of the splitting function depends
on one single parameter $\Lambda$, the radius of the form factor,
which is related by
the chiral Ward identity to
the constituent quark mass $M_q$. The large $Q^2$ tail
depends also on the axial coupling
$g_\pi$, which is expected
to be
$g_\pi\lsim{1\over2}$.

It may now be checked  explicitly
that ${\cal P}_{qq}$  can be
computed in the LLA, in that the cross
section is dominated by a collinear singularity where the coupling
is effectively pointlike. The anomalous dimensions
are  sizable for values of $Q^2$ up to 5--10~GeV$^2$, while
 at both very small
$Q^2\sim 0.05$~GeV$^2$, and at large $Q^2 > 10$~GeV$^2$, all the
anomalous dimensions flatten, ensuring
smooth connection to a valence quark picture in the infrared and to
the usual perturbative behaviour in the ultraviolet. As expected,
${\cal P}^{ND}$ is larger than
${\cal P}^{ND}$ by
roughly a factor two , thus leading to significant nonsinglet evolution.

The scale dependence of $S_G$ can now be determined: the result
obtained  assuming $S_G$ to take the ``quark model'' value
$S_G=1/3$ at a reference scale $Q_0\sim 200$~MeV is shown in Fig.2 as
a function of the two parameters $M_q$ and $g_\pi$, and compared to
the experimental result. Very good agreement is obtained with
reasonable values of the parameters. This means that
the full nonsinglet sea is generated dynamically: therefore, not only its
first moment but its full shape can now be predicted\cite{bbfg}.

To this purpose,  we need  a
model for the starting valence and its evolution at low scale. Once
this is specified\cite{barone},  the nonsinglet structure function is fully
determined at all $x$ and $Q^2$, because the bulk of the  low--$Q^2$
nonperturbative effects is  expected to be encapsulated by the
anomalous evolution equations (\ref{qev}--\ref{piev}).
The
scale dependence is driven by two competing mechanisms: gluon
radiation (Fig.1a) which softens the valence quark distribution,
and bound state emission (Fig.1c) which leads to  to a net increase
in the number of sea $q$--$\bar q$ pairs.
Thus, the peak of the nonsinglet
distribution as a function of $x$ remains approximately fixed at its
starting value $x\sim {1\over 3}$, rather than shifting towards smaller $x$
as it happens in the singlet case.
However, the overall decrease of the structure function is stronger
than found with ordinary perturbative evolution.

The structure function determined thus is compared to
the data\footnote{Notice that the data have been recently
revised;
the new
 data agree better with our  prediction \cite{bbfg} than those
\cite{oldNMC} to which we originally \cite{bbfg} compared it.}
and to the result of ordinary perturbative evolution in Fig.3. It is
apparent
that  the effects of anomalous evolution are large for all values of
$x$. The (unphysical) effect of switching gluon emission off is also shown.
In the large-$x$ region, where the NMC data have large errors, an
alternative comparison can be made with more precise data at a larger
value of $Q^2$ \cite{BCDMS}.
The stability of these results upon variation of the parameters
has been checked explicitly \cite{bbfg}.

\bigskip

Drell-Yan scattering can provide a direct determination of
\be
\label{ratdef}
{R(x)\equiv{\bar d(x)-\bar u(x)\over \bar d(x)+\bar u(x)}.}
\ee
Predicting the value of $R(x)$ in the anomalous
evolution scenario would require a computation of singlet evolution as
well; we may however estimate it by assuming the bulk of the effect of
anomalous evolution on the singlet to be just the production of an
asymmetric sea.
The result (Fig.4) has the generic feature of
saturating to a constant when $x\gsim 0.2$ because at large $x$ the
symmetric sea vanishes more rapidly than the anomalously generated
asymmetric one (as discussed above), thus only the latter
contribution to $R$ survives, which is just
equal to a constant group-theoretic
factor.
Our result is in excellent agreement  with the  data
\cite{E772,NA51}, which however cannot
exclude some of the alternative models; a measurement of $R$
Eq.(\ref{ratdef})  for several values of $x$ would be required to
discriminate between them.

The scale dependence of the full nucleon structure function $F^p_2$ is
dominated by nonsinglet evolution if $x\gsim 0.3$ because the gluon
distribution falls very rapidly at large $x$. Thus, anomalous
evolution will affect the scale dependence of $F_2^p$ at large $x$ and
intermediate values of $Q^2$. We may estimate this effect
 by again assuming that the bulk of the
anomalous contribution to both singlet and nonsinglet is just due to
the diagram of Fig.1c. Then, in this range of $x$ and
$Q^2$ $F_2^p$ will decrease less rapidly as a function of $\log Q^2$
then it would if anomalous evolution were neglected, essentially
because of the anomalous production of $q$--$ \bar q$ pairs discusse
above, which counters the softening of the structure function (Fig.5a). Of
course, at large $Q^2$ the usual results are regained.

A decrease of $|d\ln F_2^p/d \ln Q^2| $ of the same size
is found if $\alpha_s(M_z)$ is reduced by roughly 10 \% (Fig5b, from
Ref.\cite{vir}).
Because
the value of $\alpha_s$ is extracted \cite{vir} from $F_2$ by
fitting its scale dependence while neglecting anomalous evolution, if
the effect is indeed present the value of $\alpha_s$ is
systematically underestimated. Of course the relevant
fits include also the small-$Q^2$ region, where higher-twist effects are
dominant, and the large-$Q^2$ region, were anomalous evolution is
negligible; a new global fit is thus required to pin down the size of
the effect. The sign of the effect and order of magnitude, however,
appear to be just what required in order to bring this
determination  of
$\alpha_s$ in line with the LEP value.

\bigskip
{\bf Acknowledgements}: Part of the work described here was done in
collaboration with V.~Barone and M.~Genovese. We thank G.~Altarelli
for pointing out the relevance of our results to the determination of
$\alpha_s$, and M.~Arneodo and C.~Quigg for discussions.
\vfill
\eject

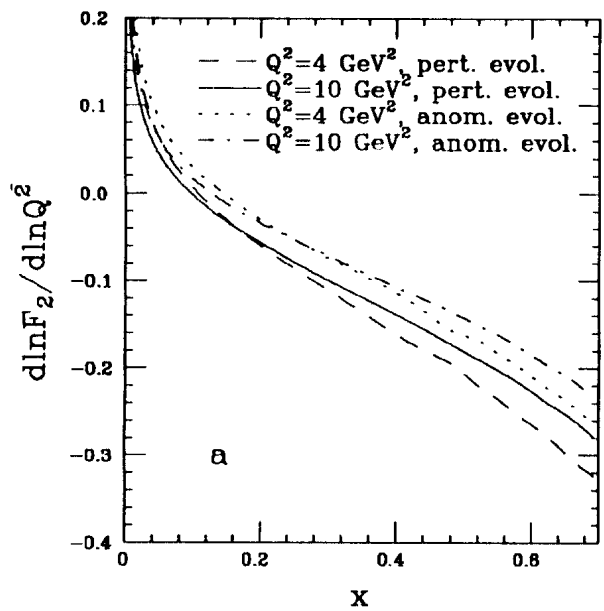
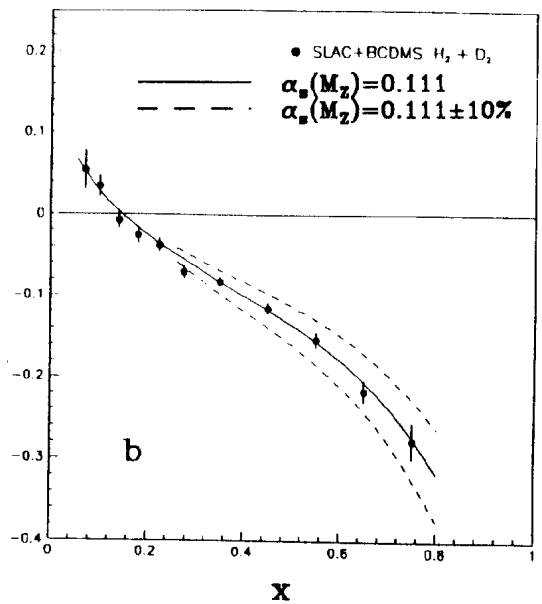
\end{document}